# Analytic evaluation of Coulomb integrals for one, two and three-electron distance operators, $R_{C1}^{-n}R_{D1}^{-m}$, $R_{C1}^{-n}r_{12}^{-m}$ and $r_{12}^{-n}r_{13}^{-m}$ with n, m=0,1,2

Sandor Kristyan

Research Centre for Natural Sciences, Hungarian Academy of Sciences,
Institute of Materials and Environmental Chemistry
Magyar tudósok körútja 2, Budapest H-1117, Hungary, kristyan.sandor@ttk.mta.hu

**Abstract**

The state of the art for integral evaluation is that analytical solutions to integrals are far more useful than numerical solutions. We evaluate certain integrals analytically that are necessary in some approaches in quantum chemistry. In the title, where R stands for nucleus-electron and r for electron-electron distances, the (n,m)=(0,0) case is trivial, the (n,m)=(1,0) and (0,1) cases are well known, fundamental milestone in integration and widely used in computation chemistry, as well as based on Laplace transformation with integrand $\exp(-a^2t^2)$. The rest of the cases are new and need the other Laplace transformation with integrand $\exp(-a^2t)$ also, as well as the necessity of a two dimensional version of Boys function comes up in case. These analytic expressions (up to Gaussian function integrand) are useful for manipulation with higher moments of inter-electronic distances, for example in correlation calculations.

**Keywords**
Analytic evaluation of Coulomb integrals for one, two and three-electron operators,
Higher moment Coulomb operators $R_{C1}^{-n}R_{D1}^{-m}$, $R_{C1}^{-n}r_{12}^{-m}$ and $r_{12}^{-n}r_{13}^{-m}$ with n, m=0,1,2

## 1. Introduction

For one-electron density, $\rho$, the $\int\rho(1)R_{C1}^{-n}R_{D1}^{-m}d\mathbf{r}_1$, $\int\rho(1)\rho(2)R_{C1}^{-n}r_{12}^{-m}d\mathbf{r}_1d\mathbf{r}_2$ and $\int\rho(1)\rho(2)\rho(3)r_{12}^{-n}r_{13}^{-m}d\mathbf{r}_1d\mathbf{r}_2d\mathbf{r}_3$ integrals own the properties: 1., The first two are not, but the last one is invariant to an interchange of n and m, 2., Trivial case: if (n,m)=(0,0) then these reduce to $(\int\rho(1)d\mathbf{r}_1)^i = N^i$ for i=1,2 and 3, resp., where N is the number of electrons in the system, 3., The well known case: if (n,m)=(1,0) or (0,1) then these are the $\int\rho(1)R_{C1}^{-1}d\mathbf{r}_1$ and $\int\rho(1)\rho(2)r_{12}^{-1}d\mathbf{r}_1d\mathbf{r}_2$ integrals, 4., The rest values for (n,m) need analytical evaluation, and constitute the topic of this work. Furthermore, integrals such as $\int\rho(1)\rho(2)\rho(3)\rho(4)r_{12}^{-n}r_{34}^{-m}d\mathbf{r}_1d\mathbf{r}_2d\mathbf{r}_3d\mathbf{r}_4 = (\int\rho(1)\rho(2)r_{12}^{-n}d\mathbf{r}_1d\mathbf{r}_2)(\int\rho(1)\rho(2)r_{12}^{-m}d\mathbf{r}_1d\mathbf{r}_2)$, etc. break up to simpler elements and fall into the cases discussed. As n and m scan the values 0,1,2, these integrals are one, two or three-electron Coulomb integrals, and next in the Introduction we list the new cases in order of their physical importance, as well as expressions will be derived not for $\rho$, but for primitive Gaussians, $G_{Ai}$, (because the real or any physically realistic model $\rho(1) \geq 0$ can be well approximated as linear combination of a well chosen set $\{G_{Ai}\}$). The standard definitions, - especially for distances - are listed at the end, which helps to guide the reader.

## 1.a. Two and three-electron Coulomb integrals for electron-electron interactions with $r_{12}^{-n} r_{13}^{-m}$

The Coulomb interaction between two charges in classical physics is $Q_1 Q_2 r_{12}^{-n}$, and is one of the most important fundamental interactions in nature. The power "n" has the rigorous value 2 describing the force, while as a consequence, the n=1 yields the energy. In quantum physics and computation chemistry [1-3], the HF-SCF and post-HF-SCF, DFT as well as CI main theories, all based on Slater determinants, provide approximate, but formally similar expressions [4-5] for electron-electron interactions, wherein the exact theory says that the Coulomb interaction energy is represented by the two-electron energy operator $r_{12}^{-1}$.

Using GTO functions, which is

$$G_{Ai}(a, nx, ny, nz) \equiv (x_i - R_{Ax})^{nx} (y_i - R_{Ay})^{ny} (z_i - R_{Az})^{nz} \exp(-a|\mathbf{r}_i - \mathbf{R}_A|^2) \quad (Eq.1)$$

with a>0 and nx, ny, nz ≥0 benefiting its important property such as $G_{Ai}(a,nx,ny,nz) G_{Bi}(b,mx,my,mz)$ is also (a sum of) GTO, the Coulomb interaction energy for molecular systems is expressed finally with the linear combination of the famous integral

$$\int G_{A1} G_{B2} r_{12}^{-1} d\mathbf{r}_1 d\mathbf{r}_2. \quad (Eq.2)$$

(In Eq.1 we use double letters for polarization powers i.e., nx, ny and nz to avoid "indice in indice", nx=0,1,2,… are the s, p, d-like orbitals, etc.. Especially in DFT, one ends up with more complex correction terms than Eq.2, but for main terms the seed is Eq.2.) The analytic evaluation [1, 6] of the integral in Eq.2 has been fundamental and a mile stone in the history of computation chemistry.

In view of the general and extreme power of series expansion (trigonometric Fourier, polynomial Taylor, Pade, etc.) in numerical calculations, however, practically the

$$\int G_{A1} G_{B2} r_{12}^{-2} d\mathbf{r}_1 d\mathbf{r}_2 \quad \text{as well as} \quad \int G_{A1} G_{B2} G_{C3} r_{12}^{-n} r_{13}^{-m} d\mathbf{r}_1 d\mathbf{r}_2 d\mathbf{r}_3 \quad (Eq.3)$$

with n,m=1,2 important terms have come up in computation chemistry, what we can call higher moments with respect to inter-electronic distances $r_{ij}$, though their analytical evaluations have not been provided yet. To understand why these terms can have importance in computation, we recall the KS-DFT [or its origin, HF-SCF] formalism for electron-electron repulsion energy approximation with one-electron density as
$V_{ee} \approx (1/2) \int \rho(\mathbf{r}_1) \rho(\mathbf{r}_2) r_{12}^{-1} d\mathbf{r}_1 d\mathbf{r}_2$ [or with normalized singe Slater determinant $S_0$ as
$V_{ee} \approx (N(N-1)/2) \int S_0^* S_0 r_{12}^{-1} ds_1 d\mathbf{r}_1 \dots ds_N d\mathbf{r}_N$], and the corresponding operator formalism, wherein $\rho$ is the sum of square of KS-MO's [or analogously with HF-SCF MO's], and the KS or HF-SCF MO's are the LCAO with GTO: These two (about the same, but not exactly the same) integrals for $V_{ee}$, which is one of the basic ideas in KS [or known as 2J(Coulomb integral)-K(exchange integral) in HF-SCF] formalism, account for almost the 99% of the Coulomb interaction energy (aside from basis set error and model error in $\rho$ [or S] itself), although the rest (called exchange-correlation energy, $E_{xc}$, for which e.g. the very weak $B_{Dirac} \int \rho^{4/3} d\mathbf{r}_1$ holds [or correlation energy, $E_{corr}$, for which e.g. the MP theory, the one of the earliest ones, has provided very remarkable approximation]) must also be approximated to reach chemical accuracy (1 kcal/mol). Among many-many ideas in the, as yet not-completely worked out, though very advanced correlation calculation, higher moments of inter-electronic distances, indicated in the title or Eq.3 have come up as candidate terms in the estimation.

We also mention, that instead of manipulating with the power of $r_{ij}$ in Eq.2, like in Eq.3, another algebraic way to use terms in correlation calculation is the moment expansion of $\rho$, as e.g. for the rough local moment expansion $V_{ee} \approx 2^{-1/3}(N-1)^{2/3}\int\rho(\mathbf{r}_1)^{4/3}d\mathbf{r}_1$ for main term itself, see review by Kristyan in ref. [7]. However, local moment expansion methods face to the problem of very slow convergence [8]. The weakness in it is the local operator vs. non-local operator, and it seems that a key to improve the existing Coulomb energy approximations in this directions is the use of e.g.

$$\{ \int[\rho(\mathbf{r}_1)]^p[\rho(\mathbf{r}_2)]^q r_{12}^{-1} d\mathbf{r}_1 d\mathbf{r}_2 \}^t \qquad (Eq.4)$$

non-local moment expansion for correlation effects (even for all kinds of approximation just mentioned above for $V_{ee}$ with a bit different parametrization for each), which is not considered yet in this literature [7]. A convenient immediate property of Eq.4 is that with GTO functions analytical evaluation is possible without any extra expressions if p and q are integer, since again, product of Gaussians is also Gaussians and analytical evaluation of Eq.2 is well known. (Also, Eq.4 is symmetric with the interchange of p and q, as well as proper choice of $t \neq 1$, Eq.4 can be made "scaling correct" depending on the p and the q, a useful property [2, 7, 9].) Taking things even a step further, one can combine Eqs.3 and 4 in relation to p, q, n and m and test the possible benefit in a correlation calculation.

Integrals in Eq.3 belong mathematically to the so called "explicitly correlated R12 theories of electron correlation", which bypass the slow convergence of conventional methods [1-2] by augmenting the traditional orbital expansions with a small number of terms that depend explicitly on the inter-electronic distance $r_{12}$. However, only approximate expressions are available for evaluation, for example, the equation, numbered 52 in ref. [10], suggests for the second one in Eq.3 that

$$\langle ijm|r_{12}^{-1}r_{13}^{-1}|kml\rangle \approx \Sigma_p \langle ij|r_{12}^{-1}|pm\rangle\langle pm|r_{12}^{-1}|kl\rangle, \qquad (Eq.5)$$

where the bracket notation [1-2] is used along without reducing product Gaussians to single Gaussians, as well as the GTO basis set {p} for expansion has to be a "good quality" for adequate approximation.

## 1.b. One-electron Coulomb integrals for nuclear-electron interactions with $R_{C1}^{-n}$

After mentioning a possible way above to correct for $V_{ee}$, wherein, for example, the HF-SCF or KS level $\rho$ is expanded with a linear combination of Gaussians one can use a similar kind of power expansion for $V_{ne}$, that is, the term

$$\int G_{A1} R_{C1}^{-2} d\mathbf{r}_1 \qquad (Eq.6)$$

similarly expands the opportunities to correct for $V_{ne}$. Recall [7] that (unlike the aforementioned HF-SCF and KS approximate expressions for $V_{ee}$, the) $V_{ne} = -\Sigma_{C=1,\ldots,M} Z_C \int \rho(\mathbf{r}_1) R_{C1}^{-1} d\mathbf{r}_1$ is an exact equation for $V_{ne}$, the only error entering is that not the exact but HF-SCF, KS, etc. approximations are used for $\rho$ in practice. By this reason the terms in Eq.6 has less importance to correct for $V_{ne}$ than Eq.3 to correct for $V_{ee}$. However, Eq.6 is still mathematically important, because it is a prerequisite (see below) to evaluate Eq.3 analytically, the aim and topic of this work. Analog expression of Eq.4 in relation to Eq.6 is the

$$\{ \int \rho^p R_{C1}^{-n} d\mathbf{r}_1 \}^t \qquad (Eq.7)$$

which can be discussed analogously.

Furthermore, if derivatives appear, such as $\int(\partial\rho(\mathbf{r}_1)/\partial x_1)^p R_{C1}^{-n} d\mathbf{r}_1$, $\int(\partial\rho(\mathbf{r}_1)/\partial x_1)^p \rho(\mathbf{r}_2)^q r_{12}^{-n} d\mathbf{r}_1 d\mathbf{r}_2$ or many other algebraic possibilities (recall that derivatives of $\rho$ are used frequently even by empirical reasons in DFT [3, 11], e.g. in the generalized gradient approximations), and $\rho$ is given as linear combination of Gaussians, analytical evaluation of Eq.3 and Eq.6 are fundamental building blocks for analytical integral evaluation, since not only the products, but the derivatives of Gaussians in Eq.1 are Gaussians.

**1.c More general one-electron and the mixed case two-electron Coulomb integrals with $R_{C1}^{-n}R_{D1}^{-m}$ and $R_{C1}^{-n}r_{12}^{-m}$, respectively**

These cases come up not only mathematically after the above cases, but in computation for electronic structures as well. Not going into too much details, we outline one way only as example: Applying the Hamiltonian twice for the ground state wave function simply yields $H^2\Psi_0 = E_{0,electr}H\Psi_0 = E_{0,electr}^2\Psi_0$, or $\langle\Psi_0|H^2|\Psi_0\rangle = E_{0,electr}^2$. The $H^2$ preserves the linearity and hermetic property from operator H, and if e.g. HF-SCF single determinant $S_0$ approximates $\Psi_0$ via variation principle from $\langle S_0|H|S_0\rangle$, the approximation $(\langle S_0|H^2|S_0\rangle)^{1/2} \approx E_{0,electr}$ is better than $\langle S_0|H|S_0\rangle \approx E_{0,electr}$, coming from basic linear algebraic properties of linear operators for the ground state. However, $H^2$ yields very hectic terms, the $H_{ne}^2$, $H_{ne}H_{ee}$ and $H_{ee}^2$ products show up, for example, yielding Coulomb operators belonging to the types in the title. Using $\langle S_0|H^2 S_0\rangle = \langle HS_0|HS_0\rangle$, the right side keeps the algorithm away from operators like $\nabla_1^2 r_{12}^{-1}$ at least.

**2.a One-electron spherical Coulomb integral for $R_{C1}^{-2}$**

Now $R_{C1} \equiv |\mathbf{R_C}-\mathbf{r_1}|$ and $R_{P1} \equiv |\mathbf{R_P}-\mathbf{r_1}|$, and we evaluate the one-electron spherical Coulomb integral for $G_{P1}(p,0,0,0) = \exp(-pR_{P1}^2)$ in Eq.1 analytically, i.e. the

$$V_{P,C}^{(n)} \equiv \int_{(R3)} \exp(-pR_{P1}^2)R_{C1}^{-n} d\mathbf{r_1}, \qquad (Eq.8)$$

for which n=1 is well known and 2 is a new expression below. The idea comes from the Laplace transformation for n= 1 and 2 respectively as

$$R_{C1}^{-1} = \pi^{-1/2} \int_{(-\infty,\infty)} \exp(-R_{C1}^2 t^2)dt, \qquad (Eq.9a)$$
$$R_{C1}^{-2} = \int_{(-\infty,0)} \exp(R_{C1}^2 t)dt = \int_{(0,\infty)} \exp(-R_{C1}^2 t)dt, \qquad (Eq.9b)$$

wherein notice the two different domains for integration. In this way (using Appendixes 1-2 after the e.g. middle part in Eq.9b) the $V_{P,C}^{(2)} = \int_{(-\infty,0)}\int_{(R3)}\exp(-p R_{P1}^2)\exp(R_{C1}^2 t)d\mathbf{r_1}dt = \int_{(-\infty,0)}\int_{(R3)}\exp(pt(p-t)^{-1}R_{CP}^2)\exp((t-p)R_{S1}^2)d\mathbf{r_1}dt = \int_{(-\infty,0)}(\pi/(p-t))^{3/2} \exp(pt(p-t)^{-1}R_{CP}^2)dt$. Using u:=t/(p-t) changes the domain t in (-∞,0) → u in (-1,0),
$V_{P,C}^{(2)} = \pi^{3/2}p^{-1/2}\int_{(-1,0)} (u+1)^{-1/2}\exp(p R_{CP}^2 u)du$, and using w:= $(u+1)^{1/2}$ changes the domain u in (-1,0) → w in (0,1) and yields

$$V_{P,C}^{(2)} = (2\pi^{3/2}/p^{1/2}) \int_{(0,1)} \exp(p R_{CP}^2 (w^2-1))dw = (2\pi^{3/2}/p^{1/2})e^{-v}F_0(-v), \qquad (Eq.10)$$

where $F_0(v)$ is Boys function with $v \equiv pR_{CP}^2$. For Eq.10 the immediate minor/major values come from $1 \le \exp(pR_{CP}^2 w^2) \le \exp(v \equiv pR_{CP}^2)$ if $0 \le w \le 1$ as

$$0 < \exp(-v) < [p^{1/2}/(2\pi^{3/2})]V_{P,C}^{(2)} < 1, \qquad (Eq.11)$$

and for a comparison, we recall the well known expression for n=1

$$V_{P,C}^{(1)} = (2\pi/p)\int_{(0,1)} \exp(-pR_{CP}^2 w^2)dw = (2\pi/p)F_0(v) \qquad (Eq.12)$$

with immediate minor/major values

$$0 < \exp(-v) < [p/(2\pi)]V_{P,C}^{(1)} < 1. \qquad (Eq.13)$$

Note that point $\mathbf{R_S}$ can be calculated by the m=2 case in Appendix 2, but its particular value drops, because integral value in Appendix 1 is invariant by shifting a Gaussian in R3 space. Eqs.11 and 13 tell that up to normalization factor with p, the $V_{P,C}^{(1)}$ and $V_{P,C}^{(2)}$ are in same range, roughly in (0,1). The ratio of the two is easily obtained when $R_{CP}=0$, then the integrands become unity, and

$$V_{P,C}^{(2)}(R_{CP}=0)/ V_{P,C}^{(1)}(R_{CP}=0) = (2\pi^{3/2}/p^{1/2})/(2\pi/p) = (\pi p)^{1/2} \qquad (Eq.14)$$

as well as for n=1 and 2 the lim $V_{P,C}^{(n)}$=0 if $R_{CP} \to \infty$.

Note that the integral is the type $\int \exp(-w^2)dw$ in Eq.12, a frequent expression coming up in physics, but contrary, the $\int \exp(w^2)dw$ has come up in Eq.10. The latter is infinite on domain $(0,\infty)$, otherwise similar algebraic blocks have come up in Eqs.8-13 for n=1 vs. 2, which is not surprising; but, the evaluation of $F_0(v)$ differs significantly from $F_0(-v)$. Integration in Eq.12 can be related to the "erf" function (i.e. for $F_0(v>0)$) in a calculation which is standard in programming, but lacks analytical expression, as well as the "erf" is inbuilt function in program languages like FORTRAN. However, integration in Eq.10 cannot be related to any inbuilt function like "erf", but its evaluation numerically belongs to standard devices, mainly because the integrand is a simple monotonic elementary function.

Note that, 1., The algebraic keys are in Eq.9 and Appendix 2 to evaluate Eq.8 analytically - up to Gaussian function $\exp(\pm w^2)$ in the integrand. If not GTO but STO is used in Eq.1, i.e. not $R_{P1}^2$ but $R_{P1}$ shows up in the power of Eq.8, the evaluation for the corresponding integral in Eq.8 is far more difficult, stemming from the fact that the convenient device in Appendix 2 cannot be used. A simple escape route is to use the approximation $\exp(-pR_{P1}) \approx \Sigma_{(i)} c_i G_{P1}(a_i,0,0,0)$, which is well known in molecular structure calculations, see the idea of STO-3G basis sets and higher levels in which one does not even need many terms in the summation but, in fact in this way, one loses the desired complete analytical evaluation for the original integral $\int_{(R3)} \exp(-pR_{P1}) R_{C1}^{-n} d\mathbf{r}_1$. 2., In Eq.9 the power correspondence in the integrand and integral value for n=1 vs. 2 is $R_{C1}^{-1} \leftrightarrow R_{C1}^{2}$ vs. $R_{C1}^{-2} \leftrightarrow R_{C1}^{2}$, what is the seed of trick for analytical evaluation, and may indicates the way for further generalizations. 3., Fast, accurate and fully numerical integration for one-electron Coulomb integrals in Eq.8 is available for any n≥1 integer and non-integer values of n, the general numerical integral scheme is widely used in DFT correlation calculations based on Voronoi polygons, Lebedev spherical integration and Becke' scheme in R3 (see brief review of related citations in ref.[4]). However, this numerical process is definitely not applicable for two and three-electron Coulomb integrals in R6 or R9, respectively because it is extremely slow in computation; the reason being that the at least K=1000 points for numerical integration becomes $K^2$ or $K^3$, respectively, that is, the computation time is K or $K^2$ times longer, respectively.

**2.b One-electron non-spherical Coulomb integral for $R_{C1}^{-2}$**

If the more general $G_{P1}(p,nx,ny,nz)$ is used, Eq.8 generates the analytical evaluation as a seed, and no further trick needed than Eq.9, the only formula necessary is how to shift the center of polynomials (Appendix 3). We use the notations $^{full}V_{P,C}^{(n)}$ and $V_{P,C}^{(n)}$, the former stands for any (spherical and non-spherical, nx+ny+nz≥0) quantum number, while the letter denotes the simplest spherical (1s-like) case, nx=ny=nz=0. Before we outline the evaluation for

$$^{full}V_{P,C}^{(2)} \equiv \int_{(R3)} G_{P1}(p,nx1,ny1,nz1) R_{C1}^{-2} d\mathbf{r}_1 , \qquad (Eq.15)$$

we should mention that Gaussians in Eq.1 are called, more precisely, "Cartesian Gaussian", and alternatively, the "Hermite Gaussians" have also been defined, see Appendix 4. All integrands in Eqs.1-3 generate Gaussians, but Hermite Gaussians are special linear combination of Cartesian Gaussians owing good recurrence and overlap relations which are not reviewed here. Appendix 2 shows that the product of two Cartesian Gaussians yields an overlap distribution, and the Gaussian product rule reduces two-center integrals to one-center integrals, (see how the $\mathbf{R}_S$ enters between Eq.9 and Eq.10). However, in the case of non-spherical (nx+ny+nz>0) Gaussians in Eq.1, large summation of Cartesian monomials is needed to evaluate the integration of non-spherical Cartesian Gaussians in Eq.15, after applying Appendices 2-3 to locate that $\mathbf{R}_S$ between $\mathbf{R}_P$ and $\mathbf{R}_C$. (Recall that the spherical nx+ny+nz=0 case in Eq.10 has only one compact term.) In practice, one expands Cartesian overlap distributions in Hermite Gaussians to evaluate $^{full}V_{P,C}^{(1)}$ analytically, and utilizes the simpler integration properties of Hermite Gaussians. Instead of recalling and using these relations, we yet use the former, since the summation is still compact; but in practice the latter is faster, tested and known for $^{full}V_{P,C}^{(1)}$. But of course, both yield the same final values.

In Eq.15 we use triple letters (nx1, etc.) which benefits in use below to distinguish between electrons 1 and 2. We apologize the reader, but from now on, lengthier equations come up, and we write out some intermediate formulas to ensure the reader that the final equations are valid. Using Eq.A.3.1 for POLY($x_1$,P,S,nx1), POLY($y_1$,P,S,ny1), POLY($z_1$,P,S,nz1) and Eq.9, the

$$^{full}V_{P,C}^{(2)} = \Sigma_1 \int_{(R3)} \int_{(-\infty,0)} (x_S-x_P)^{nx1-i1}(x_1-x_S)^{i1}(y_S-y_P)^{ny1-j1}(y_1-y_S)^{j1}(z_S-z_P)^{nz1-k1}(z_1-z_S)^{k1}$$
$$\exp(-pR_{P1}^2)\exp(tR_{C1}^2)dt d\mathbf{r}_1.$$

Using Eq.A.2.3, $\exp(-pR_{P1}^2)\exp(tR_{C1}^2) = \exp(ptR_{CP}^2/(p-t))\exp((t-p)R_{S1}^2)$, where

$x_S=(px_P-tx_C)/(p-t) \Rightarrow x_S-x_P=t(x_P-x_C)/(p-t)$ and so for x an y, so the integrand becomes
$$(t/(p-t))^{n1-m1}(x_P-x_C)^{nx1-i1}(x_1-x_S)^{i1}(y_P-y_C)^{ny1-j1}(y_1-y_S)^{j1}(z_P-z_C)^{nz1-k1}(z_1-z_S)^{k1}$$
times $\exp(ptR_{CP}^2/(p-t))\exp((t-p)R_{S1}^2)$ with short hand abbreviations (for sum and multiplication operators)

$$\Sigma_1 \equiv \Sigma_{i1=0}^{nx1} \Sigma_{j1=0}^{ny1} \Sigma_{k1=0}^{nz1} \binom{nx1}{i1}\binom{ny1}{j1}\binom{nz1}{k1} \quad \text{for even i1, j1, k1 only} \quad (\text{Eq.16a})$$
$$n1 \equiv nx1+ny1+nz1 \tag{Eq.16b}$$
$$m1 \equiv i1+j1+k1 \tag{Eq.16c}$$
$$\Gamma_1 \equiv \Gamma((i1+1)/2)\Gamma((j1+1)/2)\Gamma((k1+1)/2) \tag{Eq.16d}$$
$$D \equiv (x_P-x_C)^{nx1-i1}(y_P-y_C)^{ny1-j1}(z_P-z_C)^{nz1-k1} \tag{Eq.16e}$$

which allows for integrating out for $\mathbf{r}_1$ via Appendix 1, yielding
$$^{full}V_{P,C}^{(2)} = \Sigma_1 \Gamma_1 D \int_{(-\infty,0)} (t/(p-t))^{n1-m1} (p-t)^{-(m1+3)/2} \exp(ptR_{CP}^2/(p-t))dt.$$

Important warning is that summation in Eq.16a is for even i1, j1, k1= 0,2,4,6,… only, coming from the property of odd powers in Appendix 1. Using u:=t/(p-t) and thereafter $w^2$:=u+1 changes the domain t in $(-\infty,0) \rightarrow$ u in (-1,0) $\rightarrow$ w in (0,1), and one ends up with

$$^{full}V_{P,C}^{(2)} = 2\Sigma_1 \Gamma_1 D\, p^{-(m1+1)/2} \int_{(0,1)} (w^2-1)^{n1-m1} w^{m1} \exp(p R_{CP}^2(w^2-1))\, dw. \quad (\text{Eq.17a})$$

If n1=0, then Eq.17a reduces to Eq.10 as expected ($\Gamma(1/2)$ product provides the $\pi^{3/2}$). For example, in a full d-orbital case (nx1=ny1=nz1=2) Eq.17a has $2^3=$ terms via Eq.16a for one primitive Gaussian in Eq.15, which is not so bad, and the integrand falls to more than one terms but, by using Hermite Gaussians, the calculation is more effective. However, Eq.17a is very compact mathematically and just embedded do-loops in programming. Furthermore, since m1 is always even via Eq.16a, it yields that integrand in Eq.17a is always linear combination of $w^{2L}\exp(pR_{CP}^2(w^2-1))$ for L=0,1,2,…, i.e. Boys function can be recalled again as in Eq.10, that is, $e^{-v}F_L(-v)$ with v$\equiv$ p $R_{CP}^2$.

The expression for n=1 (in $R_{C1}^{-n}$) comes out in analogous way with the help of the known substitution $w^2$:= $t^2/(p+t^2)$, and the final integral is
$$^{full}V_{P,C}^{(1)} = 2p^{-1}\pi^{-1/2}\Sigma_1\Gamma_1 D\, p^{-m1/2}\int_{(0,1)} (-w^2)^{n1-m1}(1-w^2)^{m1/2}\exp(-p R_{CP}^2 w^2)dw. \quad (\text{Eq.17b})$$

Eq.17b reduces to Eq.12 if n1=0 in Eq.16b as expected, and since powers of $w^2$ appear, it makes the linear combination of Boys functions $F_L(v)$ with v$\equiv$ $pR_{CP}^2$. De-convolution of Boys functions from $F_L(\pm v)$ to $F_0(\pm v)$ can be found in Appendix 5. Note that D in Eq.16e dynamically provides signs.

## 2.c One-electron spherical Coulomb integral for $R_{C1}^{-n} R_{D1}^{-m}$ with n, m=1,2

As in Eq.10 vs. Eq.17, the non-spherical case is an extension of simplest spherical case with a hectic but obvious systematic summation with respect to powers (or quantum numbers) nx, ny and nz, as well as polynomials of spatial coordinates, however, the algebraic seed is Eq.10 itself. The summation is the same kind in this section and in the rest of the article too, so to save space, we derive the simplest spherical cases only.

We evaluate analytically the one-electron spherical Coulomb integral

$$V_{P,CD}^{(n,m)} \equiv \int_{(R3)} \exp(-pR_{P1}^2) R_{C1}^{-n} R_{D1}^{-m} d\mathbf{r}_1 . \qquad \text{(Eq.18)}$$

Depending on the value of (n,m), the proper one of Eq.9 must be picked. Let us take the example of (n,m)= (1,2). Using Eq.9a and e.g. the far right side in Eq.9b for D as $R_{D1}^{-2} = \int_{(0,\infty)} \exp(-R_{D1}^2 u) du$, as well as Appendixes 1-2,

$V_{P,CD}^{(1,2)} = (\pi^{-1/2}) \int_{t=(-\infty,\infty)} \int_{u=(0,\infty)} [\int_{(R3)} \exp(-gR_{W1}^2) d\mathbf{r}_1] \exp(-f/g) du dt =$
$(\pi^{-1/2}) \int_{t=(-\infty,\infty)} \int_{u=(0,\infty)} [(\pi/g)^{3/2}] \exp(-f/g) du dt$, the location of $\mathbf{R}_W$ is irrelevant again, and finally

$$V_{P,CD}^{(1,2)} = \pi \int_{t=(-\infty,\infty)} \int_{u=(0,\infty)} g^{-3/2} \exp(-f/g) du dt \qquad \text{(Eq.19a)}$$
$$g \equiv p + t^2 + u \qquad \text{(Eq.19b)}$$
$$f \equiv p t^2 R_{PC}^2 + p u R_{PD}^2 + u t^2 R_{CD}^2 . \qquad \text{(Eq.19c)}$$

Like for Eq.10 or Eq.12, by simple substitution one can end up with $\int_{(0,1)} \int_{(0,1)} (...) dt du$ integration on unit square. This integration can be done numerically, see section 3.d. The algorithm is straightforward for other cases of (n,m).

## 3. Two and three-electron spherical Coulomb integrals
## 3.a Two-electron spherical Coulomb integral for $r_{12}^{-2}$, the (n,m)=(2,0) or (0,2) case

For $G_{P1}(p,0,0,0)$ and $G_{Q2}(q,0,0,0)$ in Eq.1 we evaluate the two-electron spherical Coulomb integral analytically, i.e.:

$$V_{PQ}^{(n)} \equiv \int_{(R6)} \exp(-pR_{P1}^2) \exp(-qR_{Q2}^2) r_{12}^{-n} d\mathbf{r}_1 d\mathbf{r}_2, \qquad \text{(Eq.20)}$$

for which n=1 is well known and 2 is a new expression below. Re-indexing Eq.10 for C→2 and R→r (i.e. electron 2 takes the role of nucleus C algebraically) yields

$$V_{P,C}^{(2)} = \int_{(R3)} \exp(-pR_{P1}^2) r_{12}^{-2} d\mathbf{r}_1 = (2\pi^{3/2}/p^{1/2}) \int_{(0,1)} \exp(pR_{P2}^2 (w^2-1)) dw , \qquad \text{(Eq.21)}$$

and similarly, Eq.12 yields

$$V_{P,C}^{(1)} = \int_{(R3)} \exp(-pR_{P1}^2) r_{12}^{-1} d\mathbf{r}_1 = (2\pi/p) \int_{(0,1)} \exp(-pR_{P2}^2 w^2) dw , \qquad \text{(Eq.22)}$$

which – depending on n - integrates $\mathbf{r}_1$ out from Eq.20. Eq.21 must be used for the new one only, yielding $V_{PQ}^{(2)} = (2\pi^{3/2}/p^{1/2}) \int_{(0,1)} \int_{(R3)} \exp(pR_{P2}^2 (w^2-1) - qR_{Q2}^2) d\mathbf{r}_2 dw$. Recalling Appendixes 1-2 yields

$V_{PQ}^{(2)} = (2\pi^{3/2}/p^{1/2}) \int_{(0,1)} \exp(pq(w^2-1) R_{PQ}^2 / (p-pw^2+q)) \int_{(R3)} \exp(-(p-pw^2+q) R_{S2}^2) d\mathbf{r}_2 dw =$
$(2\pi^3/p^{1/2}) \int_{(0,1)} (p-pw^2+q)^{-3/2} \exp(pq(w^2-1) R_{PQ}^2 / (p-pw^2+q)) dw$, which yet does not show that p and q are equivalent, but they are, and the location of $\mathbf{R}_S$ is irrelevant again. At this point elementary numerical integration can be performed again, or more elegantly using $u := (1-w^2)/(p-pw^2+q)$ thereafter $w^2 := 1-u(p+q)$ changes the domain w in (0,1) → u in $(0,(p+q)^{-1})$ → w in (0,1). Finally, with $v \equiv pq R_{PQ}^2 / (p+q)$

$$V_{PQ}^{(2)} = 2\pi^3 (pq)^{-1/2} (p+q)^{-1} \int_{(0,1)} \exp(v(w^2-1)) dw = (2\pi^3 (pq)^{-1/2} (p+q)^{-1}) e^{-v} F_0(-v) , \qquad \text{(Eq.23)}$$

where $F_0(v)$ is the Boys function, and the immediate minor/major values come from $1 \leq \exp(vw^2) \leq \exp(v)$ if $0 \leq w \leq 1$ as

$$0 < \exp(-v) < [(pq)^{1/2}(p+q)/(2\pi^3)]V_{PQ}^{(2)} < 1. \quad (Eq.24)$$

For comparison, we recall the well known expression for n=1 as

$$V_{PQ}^{(1)} = (2\pi^{5/2}/(pq)) \int_{(0,c)} \exp(-pqR_{PQ}^2 w^2) dw \quad (Eq.25)$$

with $c \equiv (p+q)^{-1/2}$ in the integration domain, and it can be expressed with Boys or with "erf" functions, and the immediate minor/major values (from w:=c/0 in the integrand)

$$0 < \exp(-v) < [pq(p+q)^{1/2}/(2\pi^{5/2})]V_{PQ}^{(1)} < 1. \quad (Eq.26)$$

In Eqs.23-26 the expressions are symmetric to interchange of p and q, as expected. The ratio of the two is easily obtained when $R_{PQ}=0$, then the integrands become unity, and

$$V_{PQ}^{(2)}(R_{PQ}=0)/V_{PQ}^{(1)}(R_{PQ}=0) = (2\pi^3(pq)^{-1/2}(p+q)^{-1}/(2c\pi^{5/2}/(pq)) = (\pi pq/(p+q))^{1/2} \quad (Eq.27)$$

as well as for n=1 and 2 the $\lim V_{PQ}^{(n)}=0$ if $R_{PQ} \rightarrow \infty$.

### 3.b Two-electron spherical Coulomb integral for the mixed term $R_{C1}^{-n}r_{12}^{-m}$ with n, m=1,2

Taking n=m=1 as an example: $\int_{(R6)} \exp(-pR_{P1}^2) \exp(-qR_{Q2}^2) R_{C1}^{-1} r_{12}^{-1} d\mathbf{r}_1 d\mathbf{r}_2 =$

$\int_{(R3)} \exp(-pR_{P1}^2)[\int_{(R3)} \exp(-qR_{Q2}^2) r_{12}^{-1} d\mathbf{r}_2] R_{C1}^{-1} d\mathbf{r}_1 =$

$\int_{(R3)} \exp(-pR_{P1}^2) [(2\pi/q) \int_{(0,1)} \exp(-qR_{Q1}^2 u^2) du] R_{C1}^{-1} d\mathbf{r}_1 =$

$(2\pi/q) \int_{(0,1)} \int_{(R3)} \exp(-pR_{P1}^2 - qR_{Q1}^2 u^2) R_{C1}^{-1} d\mathbf{r}_1 du =$

$(2\pi^{1/2}/q) \int_{u=(0,1)} \int_{t=(-\infty,\infty)} \int_{(R3)} \exp(-pR_{P1}^2 - qR_{Q1}^2 u^2 - R_{C1}^2 t^2) d\mathbf{r}_1 dt du =$

$(2\pi^{1/2}/q) \int_{(0,1)} \int_{(-\infty,\infty)} [\int_{(R3)} \exp(-g R_{W1}^2) d\mathbf{r}_1] \exp(-f/g) dt du =$

$(2\pi^{1/2}/q) \int_{(0,1)} \int_{(-\infty,\infty)} [\pi/g]^{3/2} \exp(-f/g) dt du$, using Eq.31 below (for parameter m, instead of n), Eq.9a and Appendixes 1-2, as well as the location of $\mathbf{R_W}$ drops again. Finally

$\int_{(R6)} \exp(-pR_{P1}^2) \exp(-qR_{Q2}^2) R_{C1}^{-1} r_{12}^{-1} d\mathbf{r}_1 d\mathbf{r}_2 = (2\pi^2/q) \int_{u=(0,1)} \int_{t=(-\infty,\infty)} g^{-3/2} \exp(-f/g) dt du$
(Eq.28a)

$$f \equiv pqR_{PQ}^2 u^2 + pR_{PC}^2 t^2 + qR_{QC}^2 u^2 t^2 \quad (Eq.28b)$$
$$g \equiv p + qu^2 + t^2 \quad (Eq.28c)$$

wherein u and t are not equivalent. As with Eq.10 or Eq.12, the range $\int_{(-\infty,\infty)}$ for t can be converted to $\int_{(0,1)}$ by simple substitution to end up with $\int_{(0,1)} \int_{(0,1)} (...) dt du$ integration on unit square. Alternatively, $(2\pi/q) \int_{(0,1)} [\int_{(R3)} \exp(-pR_{P1}^2 - qR_{Q1}^2 u^2) R_{C1}^{-1} d\mathbf{r}_1] du =$

$(2\pi/q) \int_{(0,1)} [\int_{(R3)} \exp(-(p+qu^2)R_{W1}^2) R_{C1}^{-1} d\mathbf{r}_1] \exp(-pqR_{PQ}^2 u^2/(p+qu^2)) du =$

$(2\pi/q) \int_{(0,1)} [(2\pi/(p+qu^2)) \int_{(0,1)} \exp((p+qu^2)R_{WC}^2 t^2) dt] \exp(-pqR_{PQ}^2 u^2/(p+qu^2)) du$,

using Eq.31 below, Appendix 2 and Eq.12 for the square bracket yielding immediately $\int_{(0,1)} \int_{(0,1)} (...) dt du$, wherein $\mathbf{R_W} = (p\mathbf{R_P} + qu^2\mathbf{R_Q})/(p+qu^2)$ has a different role than above and does not drop. Finally, by using the Boys function

$\int_{(R6)} \exp(-pR_{P1}^2) \exp(-qR_{Q2}^2) R_{C1}^{-1} r_{12}^{-1} d\mathbf{r}_1 d\mathbf{r}_2 = (4\pi^2/q) \int_{(0,1)} F_0(gR_{WC}^2) g^{-1} \exp(-f/g) du$ (Eq.29a)

$$f \equiv pqR_{PQ}^2 u^2 \quad (Eq.29b)$$
$$g \equiv p + qu^2. \quad (Eq.29c)$$

Note that $R_{WC}$ in $F_0$ in Eq.29a depends on u as $gR_{WC}^2 = (p+qu^2)|\mathbf{R_W} - \mathbf{R_C}|^2 = |p\mathbf{R_P} + qu^2\mathbf{R_Q} - g\mathbf{R_C}|^2$. The $w^2 := u^2/(p+qu^2)$, changing the domain u in (0,1) → w in $(0,(p+q)^{-1/2})$, reduces the exponential part of the integrand in Eq.29a to $\exp(-pqR_{PQ}^2 w^2)$, but the algebraic complexity becomes even worse in the other terms in the integrand. Eq.29 vs. Eq.28 shows us something about the two dimensional version of the Boys function, see section 3.d, i.e. how the two dimensional integral in Eq.28 can be reduced to one dimensional, although the Boys functions still as a notation for non-analytic

integration, so Eq.29 is just an "embedding" with respect to Eq.28. Integrations in Eqs.28 and 29 can be done numerically, see section 3.d. The algorithm is straightforward for other cases of (n,m).

**3.c Three-electron spherical Coulomb integral for $r_{12}^{-n} r_{13}^{-m}$ with n,m=1,2**

For three totally different $G_{Ai}(a,0,0,0)$ in Eq.1 we evaluate the two-electron spherical Coulomb integral analytically as follows: the
$$V_{PQS}^{(n,m)} \equiv \int_{(R9)} \exp(-p R_{P1}^2)\exp(-q R_{Q2}^2)\exp(-s R_{S3}^2) r_{12}^{-n} r_{13}^{-m}\, d\mathbf{r}_1 d\mathbf{r}_2 d\mathbf{r}_3 . \quad (Eq.30)$$
Eqs.21 and 22 provide the key substitutions for integrating out with $\mathbf{r}_2$ and $\mathbf{r}_3$, because the integrand can be separated for electrons 2 and 3 as
$$\exp(-pR_{P1}^2)\, [\exp(-qR_{Q2}^2) r_{12}^{-n}]\, [\exp(-sR_{S3}^2) r_{13}^{-m}].$$
For example, for n=m=1, Eq.22 must be applied twice for $\mathbf{r}_2$ and $\mathbf{r}_3$ with index change (1,2)→(2,1) and (1,2)→(3,1), respectively, since $\mathbf{r}_1$ is the "common variable" in the denominator:
$$V_Q^{(n=1)} = \int_{(R3)} \exp(-qR_{Q2}^2) r_{12}^{-1} d\mathbf{r}_2 = (2\pi/q)\int_{(0,1)} \exp(-qR_{Q1}^2 u^2) du = (2\pi/q) F_0(qR_{Q1}^2), \quad (Eq.31)$$
$$V_S^{(m=1)} = \int_{(R3)} \exp(-sR_{S3}^2) r_{13}^{-1} d\mathbf{r}_3 = (2\pi/s)\int_{(0,1)} \exp(-sR_{S1}^2 t^2) dt = (2\pi/s) F_0(sR_{S1}^2). \quad (Eq.32)$$
Eqs.30-32 and Appendixes 1-2 yield
$$(qs/(4\pi^2)) V_{PQS}^{(1,1)} = \int_{(0,1)}\int_{(0,1)}\int_{(R3)} \exp(-pR_{P1}^2 - qR_{Q1}^2 u^2 - sR_{S1}^2 t^2)\, d\mathbf{r}_1 du dt =$$
$$\int_{(0,1)}\int_{(0,1)} [\int_{(R3)} \exp(-gR_{W1}^2) d\mathbf{r}_1] \exp(-f/g) du dt = \int_{(0,1)}\int_{(0,1)} [\pi/g]^{3/2} \exp(-f/g) du dt,$$
and finally
$$V_{PQS}^{(1,1)} = (4\pi^{7/2}/(qs)) \int_{(0,1)}\int_{(0,1)} g^{-3/2} \exp(-f/g) du dt \quad (Eq.33a)$$
$$f \equiv pq R_{PQ}^2 u^2 + ps R_{PS}^2 t^2 + qs R_{QS}^2 u^2 t^2 , \quad (Eq.33b)$$
$$g \equiv p + qu^2 + st^2 . \quad (Eq.33c)$$

The location of point $\mathbf{R}_W$ is irrelevant again in the case of 1s-like functions. This integration can be done numerically, see section 3.d, which is still more stable and more reliable than Eq.5 because the latter is basis set choice dependent and much more complex. In Eq.33 the q and s are equivalent, but they are not equivalent with the role of p, as expected, coming from the role of electron 1 vs. {2 and 3} in Eq.30. For n and/or m=2 cases not Eq.22 but Eq.21 must be applied analogously to evaluate Eq.30, the algorithm is straightforward again.

Inclusion of the Boys function can come if the expressions with t and u are not used from Eqs.31-32, but instead the far right sides with Boys functions yielding $V_{PQS}^{(1,1)} = (4\pi^2/(qs))\int_{(R3)} F_0(qR_{Q1}^2) F_0(sR_{S1}^2) \exp(-pR_{P1}^2) d\mathbf{r}_1$. For this we have not used Appendix 1 yet. As well as this, the Boys function shows up in its integrand as in Eq.29, and its argument depends on electron coordinate. At this point the aforementioned accurate DFT numerical integration [4] can be used again, since the space R9 in Eq.30 has been reduced to R3. However, to develop this further, analytically, one should use the definition of the Boys function leading the equation back to Eq.33 to be tractable.

The way to Eq.33 was to apply Eqs.31-32, then Appendixes 1-2, yielding two dimensional integral on the unit square. Another way, analogous to Eq.29 yielding one dimensional integral on the unit segment is to apply only Eq.31 and not Eq.32 or vice versa, then Appendixes 1-2, and then Eq.25: It yields
$$(q/(2\pi)) V_{PQS}^{(1,1)} = \int_{(0,1)} [\int_{(R6)} \exp(-pR_{P1}^2 - qR_{Q1}^2 u^2) \exp(-sR_{S3}^2) r_{13}^{-1} d\mathbf{r}_1 d\mathbf{r}_3] du =$$
$$\int_{(0,1)} [(2\pi^{5/2}/((p+qu^2)s)) \int_{(0,c)} \exp(-(p+qu^2) s R_{VS}^2 w^2) dw] \exp(-pqu^2 R_{PQ}^2/(p+qu^2)) du,$$
where $\mathbf{R}_V \equiv (p\mathbf{R}_P + qu^2 \mathbf{R}_Q)/(p+qu^2)$ and does not drop, since it depends on u. Finally,
$$V_{PQS}^{(1,1)} = (4\pi^{7/2}/(qs)) \int_{(0,1)} h(u)\, g^{-1} \exp(-f/g)\, du \quad (Eq.34a)$$

$$h(u) \equiv \int_{(0,c)} \exp(-g\, s\, R_{VS}^2\, w^2)\, dw \qquad \text{(Eq.34b)}$$
$$c \equiv (g+s)^{-1/2} \qquad \text{(Eq.34c)}$$
$$f \equiv pq R_{PQ}^2 u^2 \qquad \text{(Eq.34d)}$$
$$g \equiv p + q u^2 \,. \qquad \text{(Eq.34e)}$$

Eq.34 is a one dimensional integral in contrast to the two dimensional integrals in Eq.33, both yield the same value for $V_{PQS}^{(1,1)}$, of course, as well as this h(u) in Eq.34b is the pre-stage of Boys function $F_0$ as in Eq.25. Here again as in section 3.b, Eq.34 can be considered as the two dimensional version of Boys function wherein a one dimensional Boys function is in the integrand. Again, Eq.34 is only the "embedding" form of Eq.33. Section 1.c outlines a way how Coulomb operator $r_{12}^{-n} r_{13}^{-m}$ can come up; see its cardinality in Appendix 6.

**3.d The two dimensional Boys function, its pre-equation and integration**

This case comes up if a spatial coordinate appears "twice", like electron 1 in the main title of this work for n, m>0, see Eqs.19, 28 and 33. As in Eqs.10, 12, 23 or 25 the $\int_{(0,1)}(\ldots)dt$, now again by simple substitution one can end up with $\int_{(0,1)}\int_{(0,1)}(\ldots)dtdu$ integration on unit square if necessary, which can be done numerically with a simple standard device on the unit square. These two dimensional integrals can be developed further, like the one dimensional integral in Eqs.10, 12, 23, 25, see Eq.28 vs. Eq.29 and Eq.33 vs. Eq.34, as examples. It yields the extension of the one dimensional Boys function to its two dimensional version (Eqs.29 or 34), which is not worked out and analyzed yet in the literature and will be looked at in a separate work. Note the close algebraic similarity or in fact the same type in Eqs.19, 28 and 33.

If we consider the right hand side of Eq.29a or Eq.34a as a kind of two dimensional Boys function, one can see that a one dimensional Boys function appears in its integrand. We draw attention to the fact, that at the beginning, i.e. in "seed equations" Eqs.10 and 12 we obtained the one dimensional Boys function $F_0$ via the term $g^{-3/2}\exp(-f/g)$ in the integrand as a pre-equation, (recall the derivation in middle stage e.g. as $V_{P,C}^{(2)} = \pi^{3/2}\int_{(-\infty,0)} g^{-3/2}\exp(f/g)dt$ with $f \equiv pR_{CP}^2 t$ and $g \equiv p-t$), and when the two dimensional cases came up, the same term showed up in the integrand again, but instead of function set {f(t), g(t)}, the {f(u,t), g(u,t)}, see Eqs.19, 28 and Eq.33. The $g^{-3/2}\exp(f/g)$ is the core part of integrands for all cases in the main title of this work. Finer property is that, $f = f((-u)^K, (-t)^L)$ and $g = g((-u)^K, (-t)^L)$ are 2nd and 1st order polynomials, respectively, with respect to $(-u)^K$ and $(-t)^L$, where K, L = 1 or 2; wherein the middle part of Eq.9b has been used, alternatively, with the far right side of Eq.9b the -u→u and -t→t transformations should be done in this sentence. The K, L= 1 generates $\exp(w^2)$, while the 2 generates $\exp(-w^2)$ type Gaussians in the integrand.

**Conclusions**

Analytical evaluation of Coulomb one-electron integral, $\int_{(R3)} \exp(-pR_{P1}^2)R_{C1}^{-n} d\mathbf{r}_1$, has yielded $(2\pi^{3/2}/p^{1/2})e^{-v}F_0(-v)$ for n=2 in comparison to the known $(2\pi/p)F_0(v)$ for n=1, where $F_0$ is the Boys function with $v \equiv pR_{CP}^2$, and these expressions generate the formulas not only for higher quantum numbers (non-spherical or nx+ny+nz > 0 cases), but for two and three-electron Coulomb integrals as well, as indicated in the title. The equations derived help to evaluate [12] the important Coulomb integrals

$$\int \rho(1) R_{C1}^{-n} R_{D1}^{-m} d\mathbf{r}_1,$$

$$\int \rho(1)\rho(2) R_{C1}^{-n} r_{12}^{-m} d\mathbf{r}_1 d\mathbf{r}_2,$$
$$\int \rho(1)\rho(2)\rho(3) r_{12}^{-n} r_{13}^{-m} d\mathbf{r}_1 d\mathbf{r}_2 d\mathbf{r}_3$$

for n, m=0, 1, 2 in relation to powers of distances among the elements in the set of electrons and nuclei.

**Appendix 1:** For m= 1 and 2, the $\int_{(0,\infty)} x^n \exp(-ax_1^m) dx_1 = \Gamma[(n+1)/m]/(m\, a^{(n+1)/m})$ holds for a>0.

If m=2 and n=0 $\Rightarrow \int_{(R3)} \exp(-ar_1^2) d\mathbf{r}_1 = (\int_{(-\infty,\infty)} \exp(-ax_1^2) dx_1)^3 = (\pi/a)^{3/2}$.

If m=2 $\Rightarrow \int_{(-\infty,\infty)} x^n \exp(-ax_1^2) dx_1 = \Gamma[(n+1)/2]/a^{(n+1)/2}$ for even n, but zero if n is odd. The gamma function is $\Gamma[n+1] = n!$ for n=0,1,2,…, with $\Gamma[1/2] = \pi^{1/2}$ and $\Gamma[n+1/2] = 1\times 3\times 5\times\ldots(2n-1)\pi^{1/2}/2^n$ for n=1,2,… . The $\mathrm{erf}(x) \equiv 2\pi^{-1/2} \int_{(0,x)} \exp(-w^2) dw$, for which $\mathrm{erf}(\infty)=1$.

**Appendix 2**: The product of two Gaussians, $G_{J1}(p_J,0,0,0)$ with J=1,…,m=2 is another Gaussian centered somewhere on the line connecting the original Gaussians, but a more general expression for m>2 comes from the elementary
$$\Sigma_J p_J R_{J1}^2 = (\Sigma_J p_J) R_{W1}^2 + (\Sigma_J \Sigma_K p_J p_K R_{JK}^2)/(2\Sigma_J p_J) \quad \text{(Eq.A.2.1)}$$
$$\mathbf{R}_W \equiv (\Sigma_J p_J \mathbf{R}_J)/(\Sigma_J p_J) \quad \text{(Eq.A.2.2)}$$
where $\Sigma_{J\,or\,K} \equiv \Sigma_{(J\,or\,K=1\,to\,m)}$ and $R_{J1} \equiv |\mathbf{R}_J - \mathbf{r}_1|$ for $\exp(\Sigma_J c_J) = \Pi_{(J=1\,to\,m)} \exp(c_J)$, keeping in mind that $R_{JJ}=0$, and the m centers do not have to be collinear. For m=2, this reduces to
$$p R_{P1}^2 + q R_{Q1}^2 = (p+q) R_{W1}^2 + pq R_{PQ}^2/(p+q) \quad \text{(Eq.A.2.3)}$$
yielding the well known and widely used
$$G_{P1}(p,0,0,0)\, G_{Q1}(q,0,0,0) = G_{W1}(p+q,0,0,0) \exp(-pq R_{pq}^2/(p+q)) . \quad \text{(Eq.A.2.4)}$$
We also need the case m=3, which explicitly reads as
$$p R_{P1}^2 + q R_{Q1}^2 + s R_{S1}^2 = (p+q+s) R_{W1}^2 + (pq R_{PQ}^2 + ps R_{PS}^2 + qs R_{QS}^2)/(p+q+s) . \quad \text{(Eq.A.2.5)}$$
Only the $G_{W1}(p+q+s,0,0,0)$ depends on electron coordinate $\mathbf{r}_1$ in Eq.A.2.4-5, not the other multiplier, indicating that the product of Gaussians decomposes to (sum of) individual Gaussians, (s=0 reduces Eq.A.2.5 to Eq.A.2.4).

**Appendix 3:** Given a single power term polynomial at $\mathbf{R}_P$, we need to rearrange or shift it to a given point $\mathbf{R}_S$. For variable x, this rearrangement is $(x-x_P)^n = \Sigma_{i=0\,to\,n} c_i (x-x_S)^i$, which can be solved systematically and immediately for $c_i$ by the consecutive equation system obtained from the $0,1,…n^{th}$ derivative of both sides at $x := x_S$, yielding
$$\mathrm{POLY}(x,P,S,n) \equiv (x-x_P)^n = \Sigma_{i=0\,to\,n} \binom{n}{i} (x_S - x_P)^{n-i} (x-x_S)^i , \quad \text{(Eq.A.3.1)}$$
where $\binom{n}{i} = n!/(i!(n-i)!)$. If $x_S=0$, it reduces to the simpler well known binomial formula as $(x-x_P)^n = \Sigma_{i=0\,to\,n} \binom{n}{i} (-x_P)^{n-i} x^i$.

**Appendix 4:** The Hermite Gaussians are defined as
$$H_{Ai}(a,t,u,v) \equiv (\partial/\partial R_{Ax})^t (\partial/\partial R_{Ay})^u (\partial/\partial R_{Az})^v \exp(-a|\mathbf{r}_i - \mathbf{R}_A|^2) , \quad \text{(Eq.A.4.1)}$$
and $H_{Ai}(a,2,0,0) = (\partial/\partial R_{Ax})^2 \exp(-a R_{Ai}^2) = (\partial/\partial R_{Ax})[-2a(R_{Ax}-x_i) \exp(-a R_{Ai}^2)] =$
$-2a \exp(-a R_{Ai}^2) + 4a^2 (R_{Ax}-x_i)^2 \exp(-a R_{Ai}^2) = -2a G_{Ai}(a,0,0,0) + 4a^2 G_{Ai}(a,2,0,0)$ is an example that Hermite Gaussians are linear combination of Cartesian Gaussians.

**Appendix 5:** De-convolution of Boys functions from $F_L(v) \equiv \int_{(0,1)} \exp(-vt^2) t^{2L} dt$ to $F_0(v) = \int_{(0,1)} \exp(-vt^2) dt$ for v>0 and v≤0 comes from the help of partial integration ($\int f'g = [fg] - \int fg'$) on interval [0,1] with f'=$t^M$, M≠-1 and g=exp(-$vt^2$), and K:=M+2 thereafter. After elementary calculus:

$$2v\int_{(0,1)} t^K \exp(-vt^2)dt = (K-1)\int_{(0,1)} t^{K-2} \exp(-vt^2)dt - \exp(-v) \quad (Eq.A.5.1)$$

for K=0,-1,±2,±3,±4,…, i.e. any integer except 1, and v is any real number, i.e. v>0 and v≤0. (For K=1 the $2v\int_{(0,1)} t \exp(-vt^2)dt$= 1-exp(-v) by $\int g'\exp(g(t))dt=\exp(g(t))$.) In Boys functions the K=2L ≥0 is even, so K=1 is jumped, and with K:=2L+2 Eq.A.5.1 yields

$$2vF_{L+1}(v) = (2L+1)F_L(v) - \exp(-v) . \quad (Eq.A.5.2)$$

The value of L recursively goes down to zero, and the value of $F_0(v)$ is needed only at the end. The v=0 case is trivial and the v>0 is well known in the literature but, the v<0 cases are also needed for cases described in the main title of this work.

**Appendix 6:** The cardinality in the set generated by electron-electron repulsion operator $H_{ee}^2 = (\Sigma_{i=1..N}\Sigma_{j=i+1...N} r_{ij}^{-1})^2$ comes from elementary combinatorics. $H_{ee}$ contains $\binom{N}{2}$=N(N-1)/2 and $H_{ee}^2$ contains $N^2(N-1)^2/4$ terms. In relation to integration with single Slater determinant, it contains three kinds of terms: $r_{12}^{-2}$, $r_{12}^{-1}r_{13}^{-1}$ and $r_{12}^{-1}r_{34}^{-1}$ as

$$<S^*|H_{ee}^2|S> = \binom{N}{2}\{<S^*|r_{12}^{-2}|S> + 2(N-2)<S^*|r_{12}^{-1}r_{13}^{-1}|S> + \binom{N-2}{2}<S^*|r_{12}^{-1}r_{34}^{-1}|S>\} . \quad (Eq.A.6.1)$$

The control sum $\binom{N}{2}$ + 2(N-2)$\binom{N}{2}$ + $\binom{N}{2}\binom{N-2}{2}$ = $N^2(N-1)^2/4$ holds, as well as the magnitude of cardinality of individual terms on the right in Eq.A.6.1 are $N^2$, $N^3$ and $N^4$, respectively.

**Common notations, abbreviations and definitions** (watch for upper/lower cases in definitions for distances distinguishing between electrons and nuclei)

CI = configuration interactions
DFT = density functional theory
f, g = functions with different variables in integrands
$F_L(v) \equiv \int_{(0,1)} \exp(-vt^2)t^{2L}dt$, the Boys function, L=0,1,2,…
GTO = primitive Gaussian-type atomic orbital, the $G_{Ai}(a,nx,ny,nz)$ in Eq.1
H ≡ $H_\nabla + H_{ne} + H_{ee}$ = non-relativistic electronic Hamiltonian for the sum of kinetic motion, and nuclear-electron and electron-electron interactions, respectively
HF-SCF = Hartree-Fock self-consistent field
KS = Kohn-Sham
LCAO = linear combination of atomic orbitals
MO = molecular orbital
MP = Moller-Plesset
N = number of electrons in the molecular system
$Q_i$ = charge of classical particle i
R3 = 3 dimension spatial space as domain for $\int_{(R3)}...d\mathbf{r}_1 \equiv \int_{(-\infty,\infty)}\int_{(-\infty,\infty)}\int_{(-\infty,\infty)}...dx_1dy_1dz_1$
R6 = R3xR3 = 6 dimension domain for $\int_{(R3xR3)}...d\mathbf{r}_1d\mathbf{r}_2$, as well as R9=R3xR3xR3
$\mathbf{R}_A \equiv (R_{Ax}, R_{Ay}, R_{Az})$ or $(x_A, y_A, z_A)$ = 3 dimension position (spatial) vector of nucleus A
$R_{AB} \equiv |\mathbf{R}_A - \mathbf{R}_B| = ((R_{Ax}-R_{Bx})^2+(R_{Ay}-R_{By})^2+(R_{Az}-R_{Bz})^2)^{1/2}$ = nucleus-nucleus distance
$R_{Ai} \equiv |\mathbf{R}_A - \mathbf{r}_i| = ((R_{Ax}-x_i)^2+(R_{Ay}-y_i)^2+(R_{Az}-z_i)^2)^{1/2}$ = nucleus-electron distance
$\mathbf{r}_i \equiv (x_i,y_i,z_i)$ = 3 dimemsion position (spatial) vector of electron i
$r_{ij} \equiv |\mathbf{r}_i - \mathbf{r}_j| = ((x_i-x_j)^2+(y_i-y_j)^2+(z_i-z_j)^2)^{1/2}$ = electron-electron distance
$\rho(\mathbf{r}_1)$ = one-electron density, here only ground state mentioned, more precisely the $\rho_0(\mathbf{r}_1)$, the index short hand holds as $\rho(i) \equiv \rho(\mathbf{r}_i)$ for electron i=1,2,…,N

S = single Slater determinant to approximate the ground state wave function, more precisely the $S_0(s_1\mathbf{r}_1,\ldots,s_N\mathbf{r}_N)$; not to be confused with point S at $\mathbf{R}_S$
SE = non-relativistic electronic Schrödindger equation
$s_i$ = α or β spin of electron i
STO = primitive Slater-type atomic orbital, Eq.1 with change $|\mathbf{r}_i\text{-}\mathbf{R}_A|^2 \rightarrow |\mathbf{r}_i\text{-}\mathbf{R}_A|$
STO-3G = STO is approximated with linear combination of three GTO
v = function variable, its values are $pR_{CP}^2$, $pqR_{PQ}^2/(p+q)$, etc.
$V_{ee}$ = electron-electron repulsion energy in SE
$V_{ne}$ = nuclear-electron attraction energy in SE
$Z_A$ = nuclear charge of nucleus A=1,2,..M atoms in a molecular system
()*= complex conjugate

## Acknowledgments

Financial and emotional support for this research from OTKA-K 2015-115733 and 2016-119358 are kindly acknowledged.